\RequirePackage{fix-cm}
\documentclass[twocolumn]{svjour3}
\smartqed
\usepackage{amsmath, graphicx, amssymb, overcite, subfigure, mathptmx}
\journalname{}
\begin{document}

\title{Geometry of motion and nutation stability of free axisymmetric variable mass systems}

\author{Angadh Nanjangud
}

\authorrunning{A. Nanjangud}
  
\institute{A. Nanjangud \at
  University of California Davis\\
  Davis, CA 95616\\
  USA\\
  \email{a.nanjangud@surrey.ac.uk}\\
  \emph{Present address:} of A. Nanjangud\\ 
   Surrey Space Center\\
   University of Surrey\\
   Guildford GU2 7XH\\ United Kingdom\\
}

\maketitle
\begin{abstract}
In classical mechanics, the `geometry of motion' refers to a development
to visualize the motion of freely spinning bodies. In this paper, such an
approach of studying the rotational motion of axisymmetric variable mass
systems is developed. An analytic solution to the second Euler angle characterising
nutation naturally falls out of this method, without explicitly solving the nonlinear
differential equations of motion. This is used to examine the coning motion of a
free axisymmetric cylinder subject to three idealized models of mass loss and
new insight into their rotational stability is presented. It is seen that the angular
speeds for some configurations of these cylinders grow without bounds. In spite of
this phenomenon, all configurations explored here are seen to exhibit nutational
stability, a desirable property in solid rocket motors.

\end{abstract}

\section*{Nomenclature}
\begin{tabular}{@{}lcl@{}}
  $\text B$ &=& rigid base of system\\
  ${\bf b}_i (i = 1,2,3)$ &=& dextral unit vectors attached to B\\
  $\text C$ &=& rigid massless control volume\\
  $\text F$ &=& fluid phase of variable mass system\\
  ${\bf H}^*$ &=& central angular momentum of\\ &&variable mass system\\
  $h$ &=& initial half length of cylinder \\
  ${\bf I}^*$ &=& central moment of inertia of\\ &&the variable mass system\\
  $I$ &=& transverse moment of inertia scalar\\
  $J$ &=& spin moment of inertia scalar\\
  $m$ &=& instantaneous mass of the system\\
  ${\bf M}^*$ &=& resultant external moment about $S^*$ \\
  $\bf n$ &=&  unit normal directed outward from\\ &&the exit face of $\text C$\\
  ${\bf n}_i (i = f,g,h)$ &=& dextral unit vectors attached to\\ &&an inertial frame\\
  $\text O$ &=& point on $\text B$\\
  ${\bf p}$ &=& position vector from ${\text S}^*$ to $\text P$\\
  $\text P$ &=& a generic particle within $\text C$\\
  ${\bf r}$ &=& position vector from $\text O$ to $\text P$\\
  ${\bf r}^*$ &=& position vector from $\text O$ to ${\text S}^*$\\
  $R$ &=& radius of exit plane,\\ &&and initial radius of cylindrical body\\
  $\text{S}^*$ &=& mass center of system\\
  $V_0$ &=& volume of $\text C$\\
  ${\bf v}$ &=& inertial velocity of $\text P$\\
  ${\bf v}_r$ &=& velocity of $\text P$ relative to $\text B$ \\
  $z$ &=&  instantaneous half length of cylinder \\
  $z_e$ &=& shortest distance between ${\text S}^*$ \& exit\\
  \boldmath$\alpha$ &=& angular acceleration of the system\\
  \boldmath$\omega$ &=& angular velocity of the system\\
  $\rho$ &=& density of matter within $\text C$\\
\end{tabular}

\section{Introduction}
Recent examinations into the dynamics of variable mass systems can be divided into two
distinct periods; the first begins in the mid-1940's and spans the early 1970's while the
current generation of studies begins in the 1990's. The main interest
during the first stage appears to be interest in the rocket modeling problem; here, the
contributions of Rosser et al. [\citenum{rosser}] and Thomson [\citenum{thom1, thom2, thom3}]
are essential reading on the topic. Whereas Rosser et al. were
the first to present the concept of jet damping of rotating rockets, Thomson added
to the body of work on more general variable mass systems [\citenum{thom1}] and also examined
jet damping in the specific case of solid rocket motors [\citenum{thom2}]. His classic
text [\citenum{thom3}] is notable for presenting one of the first broadly accessible
discussions on attitude dynamics of rocket-type systems. The second generation of studies
was more expansive. Apart from revisiting the derivation of the equations of
motion of continuous mass varying systems [\citenum{ekewang1, ekemao1, ekeme1}], attention
has also been paid to algorithm development for simulating rigid [\citenum{djerassi}]
and flexible systems with mass variation [\citenum{banerjee, hu}], as well as analyses of
systems with discontinuous mass variation [\citenum{cveticanin1, cveticanin2}].

Yet another aspect that has been scrutinized in this period is the stability of the
rotational (or attitude) motion of spinning rigid systems with continuous
mass loss. These investigations into the stability of axisymmetric rigid
systems with mass loss [\citenum{ekewang2, ekewang3, ekemao2}] were motivated by the
anomalous coning motion seen in some solid rocket motors (SRM) [\citenum{flandro, ekejpl}].
The resulting growth in the cone angle or nutationa angle leads to a
velocity pointing error and is referred to as a coning or nutation instability. The
set of papers by Eke and Wang [\citenum{ekewang1, ekewang2, ekewang3}] began a systematic
re-examination of attitude dynamics of general variable mass systems. In the first of these
papers [\citenum{ekewang1}], the equations of motion are revisited; apart from presenting
three forms of the equation of attitude motion, they also discuss the pros and cons
of analysis with each model. In their second paper [\citenum{ekewang2}], axisymmetric
cylindrical systems with mass loss are examined; the work presented in the present
manuscript most closely relates to their work. In the last of their
papers [\citenum{ekewang3}], a more general version of the axisymmetric system
with mass loss is examined where they show that accounting for the whirl component
of internal flow in these systems does not affect the magnitude of the transverse
angular velocity. This finding informs the approach
of the current paper where a similar non-whirling flow pattern is assumed. Following up,
Eke and Mao derive the governing equations via Kane's method [\citenum{ekemao1}] and
proceed to study a composite variable mass system: a constant mass spacecraft with
mass-varying cylindrical propellant [\citenum{ekemao2}]. More recently, Ha and
Janssens [\citenum{vanderha}] investigated the CONTOUR mishap using an identical
model for mass variation as the aforementioned papers. As the CONTOUR
spacecraft exhibits slight variations in its transverse moments of inertia, an approach
to analytically determine the angular speeds of such a nearly-symmetric system with
mass loss has recently been developed and numerically validated [\citenum{ekeme2}].

A limitation of some of these studies on variable mass systems
[\citenum{ekewang2, ekemao2, ekeme2}] is that stability information is heuristically
interpreted by examining the evolution of the angular velocity. This paper 
develops an alternative graphical approach to assess the coning motion
by generating a solution to the second Euler angle. This is used to evaluate
the system's nutational stability. The approach is also referred to as the
`geometry of motion' in classical textbooks on mechanics [\citenum{likinstext}] and
spacecraft dynamics [\citenum{kaplan}]. Apart from allowing an analytic solution
to the Euler angle characterizing nutation, it also permits visualizing the motion of
a system's angular velocity vector from body-fixed and inertial frames; the document also
covers this aspect. As shown in this paper, this technique offers a more mathematically
rigorous explanation of the coning motion than one obtained from
analysing the angular speeds alone. It has previously been shown
[\citenum{ekewang2, ekemao2}] that some idealized burns of axisymmetric cylinders
can lead to unbounded growths in the transverse angular speeds. In this paper,
it is shown that these burns still damp out the coning motion in spite of
unbounded transverse speeds. A limitation of this
work and prior studies that will need to be explored in a future study is the effect
of thrust misalignment effects on the coning motion; such effects have been studied
in greater detail recently for spinning rigid systems of constant mass
[\citenum{ayoubi, martin, javorsek}].

The structure of this paper is as follows. First, the scalar equations of attitude
motion for a general axisymmetric system are formulated and analytic solutions to these
equations are obtained. Then, based on a recent
result concerning the inertial fixedness of the central angular momentum vector of
such systems [\citenum{ekeme3, diss}], the approach to visualize the rotations of
these systems from both body-fixed and inertial frames is developed. Finally,
this theory is used to evaluate the motion of axisymmetric rigid cylinders subject to three
idealized models of mass loss. From analysis, it is deduced that all of these
freely spinning cylindrical configurations are nutationally stable i.e. the coning
motion damps out or remains bounded.

\section{Equations of Motion and the Angular Velocity of an Axisymmetric Systems}
\begin{figure}
  \includegraphics[scale=1.4]{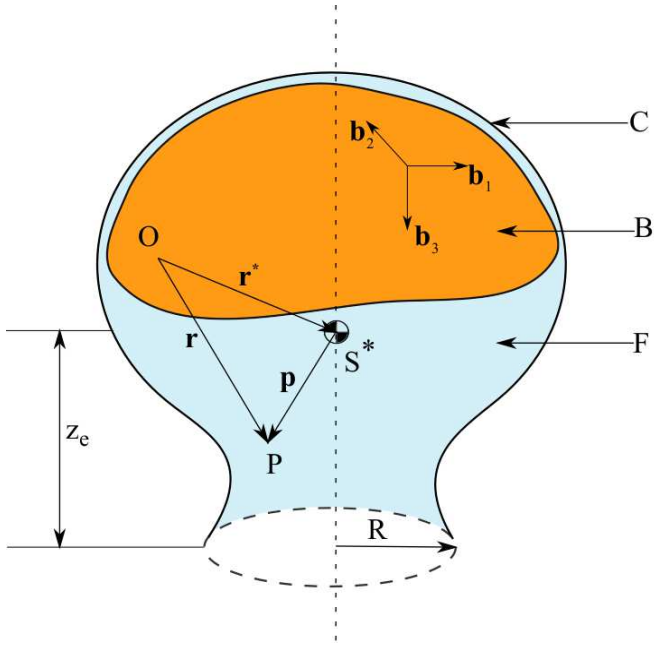}
  \caption{General variable mass system}
  \label{fig:1}
\end{figure}

Figure \ref{fig:1} is of a general variable mass system. It comprises a consumable rigid
base B and a fluid phase F. A massless shell C of constant volume $V_0$ and constant
surface area $S_0$ is rigidly attached to B. It is assumed that mass can enter or exit C
through the region represented as a dashed circle of radius $R$. At every instant, the
shell and everything within it is considered to be the system of interest.
The vector equation of attitude motion of such a system can be derived from the laws
of mechanics (such as Newton-Euler equations, Lagrange's method, etc.) and
appropriately invoking Reynold's Transport Theorem. This approach has been discussed in
several papers [\citenum{ekewang1, ekemao1, ekeme1}]. The following is the equation of
attitude motion of a general variable mass system
\begin{eqnarray}
  \label{1}
  {\bf M}^* = &&{\bf I}^* \cdot \textrm{\boldmath$\alpha$}
  + \textrm{\boldmath $\omega$} \times ({\bf I}^* \cdot \textrm{\boldmath $\omega$})
  + \frac{^Bd {\bf I}^*}{\mathrm{d}t} \cdot \textrm{\boldmath $\omega$}\nonumber\\ 
  &&+ \int_{S_0} \rho \mathbf{[p \times (\textrm{\boldmath $\omega$} \times p)]
  	(v_r \cdot {n})} \, dS \nonumber\\
  &&+ \int_{V_0} \rho \mathbf{[\textrm{\boldmath $\omega$} \times (p \times v_r)]} \, dV
  + \frac{^Bd}{\mathrm{d}t} \int_{V_0}\rho \mathbf{(p \times v_r)} \, dV\nonumber\\
  &&+ \int_{S_0} \rho \mathbf{(p \times v_r)}(\mathbf{v_r} \cdot {\mathbf n}) \, dS.
\end{eqnarray}
Several other forms of the equation of attitude motion have been derived for a variable
mass system in the literature. However, Equation (\ref{1}) is most tractable for
examining the attitude evolution as it is decoupled from the translational dynamics.
The volume integrals in Equation (\ref{1}) are typically hard to evaluate in closed-form
since the internal flow field is not generally known within a system. However,
in the case of rockets, reasonable assumptions can be made about this internal flow. It
is typically assumed that the internal flow of gases within a rocket relative to its
casing is both steady and axisymmetric. Further assuming that the flow lacks a whirl
component force the last three terms of Equation (\ref{1}) to disappear, making it
less cumbersome.  Thus, the equation of attitude motion becomes
\begin{eqnarray}
  \label{2}
  {\bf M}^* = {\bf I}^* \cdot \textrm{\boldmath $\alpha$}
  &&+ \omega \times ({\bf I^*} \cdot \textrm{\boldmath $\omega$}) 
  + \frac{^Bd {\bf I}^*}{\mathrm{d}t} \cdot \textrm{\boldmath $\omega$}\nonumber\\
  &&+ \int_{S_0} \rho \mathbf{\bigg[p \times (\textrm{\boldmath $\omega$} \times p)\bigg]
  (v_r \cdot {n})} \, dS.
\end{eqnarray}
The remainder of this paper concerns the torque-free motions of axisymmetric variable mass
systems so $\mathbf{M^* = 0}$. Note that though this non-whirling flow assumption is not
realistic, it has been shown that the magnitude of the transverse angular
rates are unaffected by such a flow assumption[\citenum{ekewang3}].
As stability information in this paper is derived from the magnitude of the angular rates,
the non-whirling flow assumption is sufficient.

Referring back to Figure \ref{fig:1}, ${\bf b}_1, {\bf b}_2, {\bf b}_3$ are a dextral set of
unit vectors affixed in B and parallel to its principal directions. The
instantaneous central inertia dyadic of this axisymmetric system is
\begin{equation}
  \label{3}
  {\bf I}^* \triangleq I \, \, {\bf b}_1 {\bf b}_1
  + I \, \, {\bf b}_2 {\bf b}_2
  + J \, \, {\bf b}_3 {\bf b}_3
\end{equation}
where $I$ and $J$ are the instantaneous central principal moments of inertia of the variable
mass system. Note that, the mass center, $S^*$ is not necessarily fixed relative to B
and is always located on the symmetry axis, parallel to ${\bf b}_3$. Further, the inertial
angular velocity of B at any instant is
\begin{equation}
  \label{4}
  \textrm{\boldmath$\omega$} \triangleq \omega_1 {\bf b}_1
  + \omega_2 {\bf b}_2
  + \omega_3 {\bf b}_3
\end{equation}
and, as a result, the inertial angular acceleration is
\begin{equation}
  \label{5}
  \textrm{\boldmath$\alpha$} = \dot\omega_1 {\bf b}_1
  +\dot \omega_2 {\bf b}_2
  +\dot \omega_3 {\bf b}_3.
\end{equation}
The first three terms on the right-hand side of Equation (\ref{2}) evaluate to
\begin{equation}
  \label{6}
  {\bf I}^* \cdot \textrm{\boldmath$\alpha$} = I \dot \omega_1{\bf  b}_1
  + I\dot \omega_2{\bf b}_2
  + J \dot \omega_3 {\bf  b}_3,
\end{equation}
\begin{equation}
  \label{7}
  \textrm{\boldmath$\omega$} \times ({\bf I}^* \cdot \textrm{\boldmath$\omega$}) =
  \bigg(J - I\bigg)\omega_2 \omega_3 {\bf b}_1
  - \bigg(J - I\bigg) \omega_1 \omega_3{\bf  b}_2,
\end{equation}
and
\begin{equation}
  \label{8}
  \frac{^Bd {\bf I}^*}{dt} \cdot \textrm{\boldmath$\omega$} = \dot I \omega_1 {\bf b}_1
  + \dot I \omega_2{\bf b}_2
  + \dot J \omega_3 {\bf b}_3,
\end{equation}
respectively.
The assumed steady symmetry of the internal fluid flow is further constrained to be uniform
along the exit plane so that ${\bf v_r \cdot n}= u = {\text constant}$, which then yields
the following expression for the surface integral in Equation (\ref{2})
\begin{eqnarray}
  \label{9}
  \int_{S_o} \rho {\bf{p \times (\textrm{\boldmath$\omega$} \times p)(v_r \cdot {n})}} \, dS = \\ \nonumber -\dot m \bigg[(z_e^2 + \frac{R^2}{4})(\omega_1 {\bf b_1} + \omega_2
  {\bf b_2}) &&+ \frac{R^2}{2}\omega_3{\bf b_3}\bigg].
\end{eqnarray}
where $\dot m$, the rate of mass loss, is given by
\begin{equation}
  \label{10}
  \dot m = -\rho_{exit} \pi R^2 u = constant.
\end{equation}
$\rho_{exit}$ is the assumed density of exhaust gases on the exit plane. 

Substituting Equations (\ref{6}), (\ref{7}), (\ref{8}), and (\ref{9}) into Equation
(\ref{2}) yields
\begin{equation}
  \label{11}
  \dot{\omega}_1 = \frac{I - J}{I} \omega_2\omega_3
  - \frac{1}{I}\bigg[\dot{I} - \dot{m}(z_e^2 + \frac{R^2}{4})\bigg] \omega_1
\end{equation}
\begin{equation}
  \label{12}
    \dot{\omega}_2 = -\frac{I - J}{I} \omega_3\omega_1
    - \frac{1}{I}\bigg[\dot I - \dot{m}(z_e^2 +
    \frac{R^2}{4})\bigg] \omega_2
\end{equation}
and
\begin{equation}
    \label{13}
    \dot{\omega}_3 = - \frac{1}{J}[\dot{J} - \dot{m}(\frac{R^2}{2})] \omega_3.
\end{equation}
Equations (\ref{11})-(\ref{13}) are the scalar equations of attitude motion for an
axisymmetric, torque-free variable mass system with axisymmetric non-whirling internal
mass flow. Equations (\ref{11}) and (\ref{12}) are commonly referred to as the equations of
transverse motion while Equation (\ref{13}) is the spin equation of motion; $\omega_1$ and
$\omega_2$ are transverse rates and $\omega_3$ is the spin rate. The attitude motion of a
variable system is characterized by solving these equations.

\subsection{Spin rate}
As the spin equation of motion is decoupled from the transverse equations of motion, an
expression for the spin rate can be easily obtained and is given by
\begin{equation}
  \label{14}
  \omega_3 = \omega_{30} \text{exp}(\Phi) 
\end{equation}
where
\begin{equation}
  \label{15}
    \Phi = 
    - \int_0^t \bigg[\bigg(\dot J - \dot m R^2/2\bigg)/J \bigg] \mathrm{d}t.
\end{equation}
If $J \triangleq m k_3^2$, where $k_3$ is the time-varying axial radius of gyration, then
$\Phi$ evaluates to
\begin{equation}
  \label{16}
  \Phi =  \int_{m(0)}^{m} \frac{(R^2/2)}{k_3^2} \frac{\mathrm{d}m}{m}
      + \mathrm{ln}\bigg(\frac{J(0)}{J}\bigg).
\end{equation}
Substituting for $\Phi$ in Equation (\ref{14}) gives the following expression for the
spin rate
\begin{equation}
  \label{17}
  \omega_3 = \omega_{30} \frac{J(0)}{J}
  \text{exp}\bigg(\int_{m(0)}^{m} \frac{(R^2/2)}{k_3^2} \frac{\mathrm{d}m}{m} \bigg)
\end{equation}
Examining Equation (\ref{17}), the spin rate may decay, stay constant, grow, or
fluctuate.  Another observation that can be added to this is that the spin rate retains the
polarity of its initial condition; if the initial condition is positive, then the spin rate is
always non-negative, and vice versa. Further analysis in this chapter assumes a positive value
for the initial spin rate. Additionally, if $k_3^2$ is assumed to be a constant in Equation
(\ref{17}), then
\begin{equation}
  \label{18}
  \omega_3 = \omega_{30} \bigg( \frac{m(0)}{m}\bigg) ^{1 - \frac{R^2}{2k_3^2}}.
\end{equation}
Equation (\ref{18}) asserts that the spin rate can not fluctuate for a system with constant
axial radius of gyration; it can only grow or decay monotonically, or stay constant. It can then
also be inferred that a time-varying axial radius of gyration can lead to fluctuations in
the spin rate. Thus, the radius of gyration has a crucial effect on the spin
rate. These comments on the spin rate are in agreement with the work of Snyder and Warner
[\citenum{snyder}], and Wang and Eke [\citenum{ekewang3}].

\subsection{Transverse rate}
The transverse angular speeds can be solved for in a variety of different ways. Defining
a complex variable $\omega_c$ as
\begin{equation}
  \label{19}
  \omega_c \triangleq \omega_1 + j \omega_2
\end{equation}
allows Equations (\ref{11}) and (\ref{12}) to be combined to yield a
differential equation linear in $\omega_c$:
\begin{equation}
  \label{20}
  \dot{\omega}_c = \Bigg\{-j \frac{I - J}{I} \omega_3
  - \frac{1}{I}\bigg[\dot{I} - \dot{m}(z_e^2 + \frac{R^2}{4})\bigg]\Bigg\} \omega_c.
\end{equation}
Integrating Equation (\ref{20}) yields
\begin{equation}
\label{21}
{\omega}_c = \omega_{co} \cdot \Gamma \cdot \text{exp} \bigg(-j\chi\bigg)
\end{equation}
where
\begin{equation}
   \label{22}
   \Gamma = \text{exp} \bigg[ -\int_0^t \frac{1}{I}\bigg[\dot{I} - \dot{m}(z_e^2 + \frac{R^2}
   {4})\bigg] \mathrm{d}t \bigg]
\end{equation}
and
\begin{equation}
  \label{23}
  \chi = \int_0^t (1 - \frac{J}{I}) \omega_3 \mathrm{d}t.
\end{equation}
A consequence of Equation (\ref{23}) is
\begin{equation}
  \label{24}
  \dot \chi = (1 - \frac{J}{I}) \omega_3
\end{equation}
where $\dot \chi$ has the units of angular speed.

From Equations (\ref{21}) and (\ref{19}), the transverse speeds are given by
\begin{equation}
  \label{25}
  \omega_1 = \Gamma (\omega_{10} \cos\chi + \omega_{20}\sin\chi)
\end{equation}
\begin{equation}
  \label{26}
  \omega_2 = \Gamma(-\omega_{10}\sin\chi + \omega_{20}\cos\chi)
\end{equation}
and are oscillatory in nature.

Further, a transverse angular velocity vector {\boldmath $\omega$}$_{t}$ which lies in
the plane made by the principal directions ${\bf b}_1$ and ${\bf b}_2$ can also be
defined as
\begin{equation}
  \label{27}
  \textrm{\boldmath $\omega$}_{t} \triangleq \omega_1 {\bf b}_1+ \omega_2 {\bf b}_2,
\end{equation}
whose magnitude is
\begin{equation}
  \label{28}
  \omega_{12} = ||\textrm{\boldmath $\omega$}_{t}||= \sqrt{\omega_1^2 + \omega_2^2} =
  \sqrt{\Gamma^2 (\omega_{10}^2 +\omega_{20}^2)} = \Gamma \omega_0
\end{equation}
where $\omega_0 \triangleq \sqrt{\omega_{10}^2 + \omega_{20}^2} = \text{constant}$. The
principal directions may be chosen such that $\omega_{10} = 0$ and $\omega_{20} =
\omega_0$. This simplifies the appearance of the solutions in Equations (\ref{25}) and
(\ref{26}) to
\begin{equation}
  \label{29}
  \omega_1 = \omega_0 \Gamma \sin\chi
\end{equation}
\begin{equation}
  \label{30}
  \omega_2 = \omega_0 \Gamma \cos\chi.
\end{equation}
Equation (\ref{27}) can then be written as
\begin{equation}
  \label{31}
  \textrm{\boldmath $\omega$}_{t} = \omega_0 \Gamma[\sin\chi {\bf b}_1+
  \cos\chi {\bf b}_2] = \omega_0 \Gamma{\bf b}_{12} = \omega_{12} {\bf b}_{12},
\end{equation}
where ${\bf b}_{12} = \sin\chi {\bf b}_1+ \cos\chi {\bf b}_2$; ${\bf b}_{12}$ is clearly
a unit vector that is orthogonal to ${\bf b}_3$. These developments concerning the
transverse angular velocity vector will be useful in developments in Section 3 on the
geometry of motion.

From examining Equations (\ref{31}) and (\ref{22}), it is evident that the spin rate has no
influence on the magnitude of the transverse angular velocity. Further, it shows that the
magnitude of the transverse angular velocity vector is a rescaling of $\omega_0$ by
$\Gamma$, a function of time that can either grow, decay, or fluctuate. ${\bf b}_{12}$ is a
unit vector that rotates in the plane of ${\bf b}_1$ and ${\bf b}_2$ at the angular speed
$\dot \chi$.  As revealed by Equation (\ref{24}), $\dot \chi$ may be positive or negative
(assuming that $\omega_3$ is always positive), depending on the relative values of $k_1$ and
$k_3$; if $k_3 > k_1$, then ${\bf b}_{12}$ rotates negatively in the body, advancing from
${\bf b}_1$ to $-{\bf b}_2$ and around; if $k_1 > k_3$, then ${\bf b}_{12}$ rotates
positively in the body, coinciding with ${\bf b}_1$ and then with ${\bf b}_2$ and so on.
As the interest of this paper is in understanding the nutational stability, no attempt
is made to obtain an analytically expression for $\chi$. Its value can be obtained
numerically from Equation (\ref{23}).

An expression for $\Gamma$ can be developed involving the system's transverse radius of
gyration, $k_1$. If $I \triangleq m k_1^2$, $\Gamma$ may be reformulated in Equation
(\ref{22}) as
\begin{equation}
    \Gamma =
    \frac{I(0)}{I}\text{exp}\bigg[ 
	       \int_{m(0)}^m \frac{\bigg[z_e^2 + \frac{R^2}{4}\bigg]}{k_1^2}
	     \frac{\mathrm{d}m}{m}\bigg]
\end{equation}
Here, $R$ is the constant radius of the exit plane
while $k_1$ and $z_e$ can vary; these are parameters that are modeling constraints.
Substituting for $\Gamma$ into Equation (\ref{28}) yields the following expression for
the magnitude of the transverse angular velocity:
\begin{equation}
  \label{33}
	\omega_{12} = \omega_{0} \bigg(\frac{I(0)}{I}\bigg)
	     \text{exp}\bigg[
	       \int_{m(0)}^m \frac{\bigg[z_e^2 + \frac{R^2}{4}\bigg]}{k_1^2}
	     \frac{\mathrm{d}m}{m}\bigg].
\end{equation}
Some of the effects of the variations in $z_e$ and $k_1$ on the transverse rate will be
studied in the subsequent passage on variable mass cylinders. However, it should be evident
that $\omega_{12}$ can grow, decay, or fluctuate but under no circumstance is it constant
for an axisymmetric variable mass system.

Finally, the inertial angular velocity of an axisymmetric variable mass system can now
be written as
\begin{equation}
  \label{32}
  \textrm{\boldmath $\omega$} = \omega_0 \Gamma{\bf b}_{12} + \omega_3 {\bf b}_3,
\end{equation}
the form of which mimics the constant-mass rigid body case, for which $\Gamma = 1$.

\section{Geometry of Torque-free Motion and Nutation Angle Solution}
Explicit description of the inertia parameters coupled with the developments regarding the
angular speeds from the previous section are sufficient to visualize the motion
of the angular velocity vector from the body-fixed frame. However, it is also
possible to visualize its motion in an inertially fixed reference frame which is the
focus of this section. These developments are then used to study the variable
mass cylinder problem in the following section.

\begin{figure}
  \centering\includegraphics[scale=1.8]{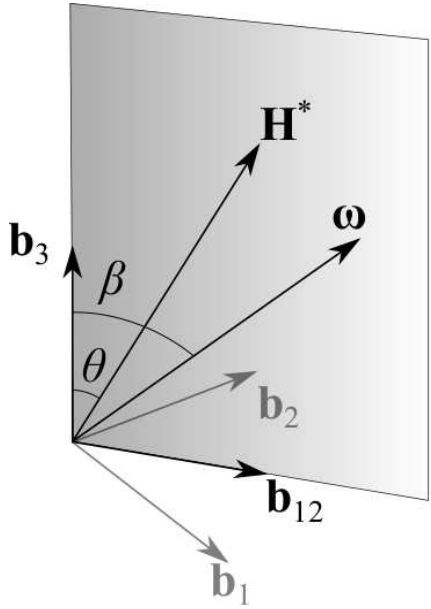}
  \caption{Geometry of motion setup of vectors}
  \label{fig:4.3}
\end{figure}

Exploiting the fact that the variable
mass system's angular momentum vector is fixed in inertial space [\citenum{ekeme3, diss}]
enables us to visually study the angular velocity evolution from an inertial frame.
The angular momentum of a general variable mass system can be written as
\begin{equation}
   \label{34}
   {\bf H}^* = \int_{V_0} \rho [{\bf p} \times[{\bf v}_r + (\textrm{\boldmath $\omega$}
   \times {\bf p}) ]\, \mathrm{d}V.
\end{equation}
Retaining the assumption of steady and axisymmetric relative internal mass flow reduces
Equation (\ref{34}) to
\begin{equation}
   \label{35}
   {\bf H}^* = \int_{V_0} \rho [{\bf p} \times (\textrm{\boldmath $\omega$}
   \times {\bf p}) ] \, \mathrm{d}V = {\bf I}^* \cdot \textrm{\boldmath $\omega$}.
\end{equation}
With the definitions of ${\bf I}^*$ and {\boldmath $\omega$} from Equations
(\ref{3}) and (\ref{4}), respectively, the component form of the angular momentum vector is
\begin{equation}
  \label{36}
  {\bf H}^* = I (\omega_1 {\bf b}_1  + \omega_2{\bf b}_2) + J \omega_3 {\bf b}_3
\end{equation}
or
\begin{equation}
  \label{37}
  {\bf H}^* = I \textrm{\boldmath $\omega$}_{t} + J \omega_3 {\bf b}_3.
\end{equation}
Using the component form of $\textrm{\boldmath $\omega$}_{t}$ from Equation (\ref{31}) in
Equation (\ref{37}) gives
\begin{equation}
  \label{38}
  {\bf H}^* = I \omega_0 \Gamma {\bf b}_{12} + J \omega_3 {\bf b}_3
\end{equation}
Equation (\ref{38}) gives an alternate form of the angular momentum of an axisymmetric
variable mass system.
Axisymmetric rigid systems (of constant or variable mass) are unique in that their angular
momentum and angular velocity vectors are always in the same plane, evident by comparing
Equations (\ref{32}) to (\ref{38}); ${\bf H}^*$ and {\boldmath$\omega$} are in the same
plane formed by ${\bf b}_{12}$ and ${\bf b}_3$ and this is shown in Figure \ref{fig:4.3}.
In this figure, $\theta$ is the angle between ${\bf H}^*$ and ${\bf b}_3$, and
$\beta$ is the angle between ${\bf b}_3$ and {\boldmath $\omega$}. These angles can
be computed from the figure as
\begin{equation}
  \label{39}
  \theta = \tan^{-1}\bigg(\frac{I \omega_{12}}{J \omega_3} \bigg)
  = \tan^{-1}\bigg(\frac{\Gamma k_1^2 \omega_{0}}{k_3^2 \omega_3} \bigg)
\end{equation}
and
\begin{equation}
  \label{40}
  \beta = \tan^{-1}\bigg(\frac{\omega_{12}}{\omega_3} \bigg) = \tan^{-1}
  \bigg(\frac{\Gamma \omega_{0}}{\omega_3} \bigg).
\end{equation}
Evaluation of these angles $\theta$ and $\beta$ along with the solutions to the angular
velocities permit a visual understanding of the motion of the body in both
inertial space and the body-fixed coordinate system.

In the case of axisymmetric rigid bodies of constant
mass, $\Gamma = 1$ and the spin rate and inertia scalars are constant. Consequently, in
Equation (\ref{39}) and (\ref{40}), both angles evaluate to constants and visualizing the
motion of $\textrm{\boldmath $\omega$}$ results in the formation of two cones:
as it rotates about ${\bf b}_3$ and another as it rotates about ${\bf n}_h$,
a unit vector parallel to the inertially fixed ${\bf H}^*$. The two cones are
appropriately named the {\it body cone} and {\it space cone}, respectively.

In the case of axisymmetric variable mass systems, angles $\theta$ and $\beta$
are typically functions of time. The angles may grow or decay as dictated by the
instantaneous values of the radii of gyration, $\Gamma$, and the spin rate.
It is also possible, in the case of the same variable mass system, for $\beta$ to exceed
$\theta$ at certain instants while it may not at certain other instants.
Figure \ref{fig:4.3} represents an instant where $\beta>\theta$, which occurs when
$k_3>k_1$. Also, $\textrm{\boldmath $\omega$}$ does not trace cones as it rotates
about ${\bf b}_3$ and ${\bf n}_h$. Thus, generic names are used for the surfaces it traces;
in this document, they are referred to as the {\it body surface} and {\it space surface}
(which reduce to the `{\it body cone}' and `{\it space cone}' for the limiting case of a
constant mass rigid body, respectively).

In the spacecraft dynamics literature, $\theta$ is referred to as
the nutation angle [\citenum{kaplan}]. In
classical mechanics, $\theta$ may also be identified as the second Euler
angle [\citenum{likinstext}] and, thus, Equation (\ref{39}) also gives the
solution to said Euler angle. Growth in nutation angles results in a coning motion
which is undesirable in spacecraft. Potential hazards of excessive nutation or coning
motion might be poor initial conditions for subsequent mission operations, or possible
loss of communications with a spacecraft due to incorrect pointing between its antenna
and the base station. One such case of excessive nutation growth was observed in
missions powered by the STAR-48 SRM [\citenum{ekejpl, meyer, flandro, cochran,
or, mingori1, mingori2}]. To mitigate this growth, an active nutation damper is
used in these missions [\citenum{webster}].

Closed form expressions to $\theta$ are available when the angular speeds are
themselves available in closed form. These closed-form expressions are invaluable
to not only evaluate stability but also for appropriate control strategy
formulation to temper any instabilities. Solutions to the angular speeds are derived in
the next section for a special class of variable mass systems: the axisymmetric variable
mass cylinder. These closed form solutions are then utilized in constructing the body
and space surfaces. The choice of cylindrical systems as an idealization is based on
their use in several prior studies on rocket flight; Snyder and Warner
[\citenum{snyder}] and Eke et al. [\citenum{ekewang2,ekemao2}] are some examples of
investigators who have incorporated cylindrical propellant grains in their studies on
the attitude motions of variable mass space vehicles.

\section{The Axisymmetric, Variable Mass Cylinder}
\begin{figure}
  \centering\includegraphics[scale=.5]{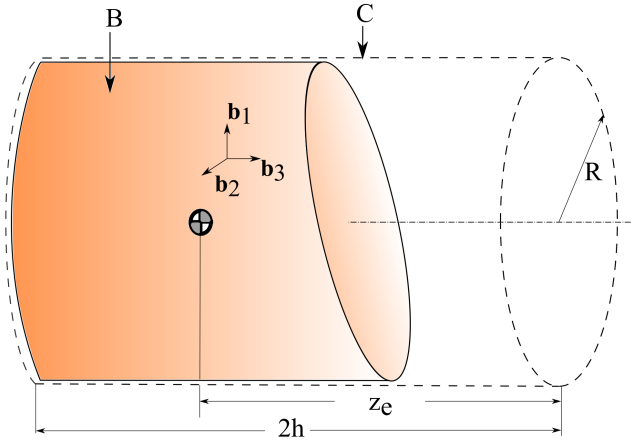}
  \caption{Variable mass cylinder}
  \label{figure:4.4}
\end{figure}
Figure {\ref{figure:4.4}} is that of a system that is initially a cylinder as represented by
the dashed lines. The mass and inertia of this system is allowed to vary with time.
As in the case of rockets, combustion is one plausible cause for these time-varying
properties. Thus, at a general instant, some parts of the system may be solid while
other parts are fluid. The solid part of the system is denoted $B$. The dotted line
represents a control region $C$ of constant volume and surface area that has the
same shape as the original cylinder and is rigidly attached to $B$. Only matter within
$C$ at any instant is considered to make up the system. Interest here is in visualizing
the motion of variable mass cylinders and understanding their nutational stability.

Equations (\ref{11})-(\ref{13}) govern the attitude motion of the variable mass
cylinder while Equations (\ref{17}), (\ref{29}), and (\ref{30}) are the solutions
to the angular speeds. Eke and Wang [\citenum{ekewang2}] identified idealized burn
scenarios of which three are examined here. They are named the uniform burn, end burn,
and radial burn. Each of these burns is explained in their respective sections.
Inertia change due to fluid mass is neglected in all burn cases. The assumed
properties of the system in creating the results presented are as follows. For the
presented results, the density of the rigid propellant is assumed to be
$1000$ kg/m$^3$. The initial angular speeds are assumed to be $\omega_0=0.2$ rad/s,
and $\omega_{30} = 0.3$ rad/s; note that in a realistic rocket scenario, the spin
rate would be an order of magnitude greater than the transverse rates. The duration
of the burn is assumed to be $100$ s unless mentioned otherwise. Other assumptions
concerning the geometry are discussed under each burn scenario.

\subsection{Uniform Burn}
A uniformly burning cylinder is one in which the mass of the system varies as a function of
time while the dimensions of the unburned portion of the cylinder are the same as that of the
original cylinder; radius and length of the cylinder are both assumed to be $1$m.  Thus, its instantaneous central moment of inertia scalars are
\begin{equation}
  \label{41}
  I = m \bigg(\frac{R^2}{4} + \frac{h^2}{3}\bigg)
\end{equation}
and
\begin{equation}
  \label{42}
  J = m\frac{R^2}{2}.
\end{equation}
This is an example of a variable mass system where the radii of gyration are constant where
$k_1^2 = R^2/4 + h^2/3$ and $k_3^2 = R^2/4$. The time derivatives of the moment of inertia
are
\begin{equation}
    \begin{aligned}
        \dot I &= \dot m(\frac{R^2}{4} + \frac{h^2}{3})\\
        \dot J &= \dot m \frac{R^2}{2}
    \end{aligned}
\end{equation}
The resulting spin and transverse rates are obtained from Equations
(\ref{17}) and (\ref{28}) as
\begin{eqnarray}
  \omega_3 &=& \omega_{30}\\
  \omega_{12} &=& \omega_0 \bigg(\frac{m}{m_0}\bigg)^{\frac{2h^2}{3k_1^2}} = \omega_0 \Gamma(t)
\end{eqnarray}
where $\Gamma(t) \triangleq (m/m_0)^{2h^2/3k_1^2}$. Prior to the start of the burn $m = m_0$ so
$\Gamma = 1$. As the burn proceeds $\Gamma$ decreases with time but is always non-negative.
Thus, the transverse angular rate also decreases with time for a uniformly burning cylinder.
Then Equation (\ref{39}) gives the nutation angle for this system
\begin{equation}
  \label{43}
  \theta(t) = \tan^{-1}( K \Gamma(t) ),
\end{equation}
where $K = k_1^2 \omega_0/k_3^2 \omega_{30}$. Since $K$ is constant and $\Gamma(t)$ is
decreasing with time, the nutation angle is also a decreasing parameter with time.
This reduction in the transverse oscillations of a variable mass system by the exhaust
gas is referred to as jet damping of a rocket.

\begin{figure}
  \centering\includegraphics[scale=0.8]{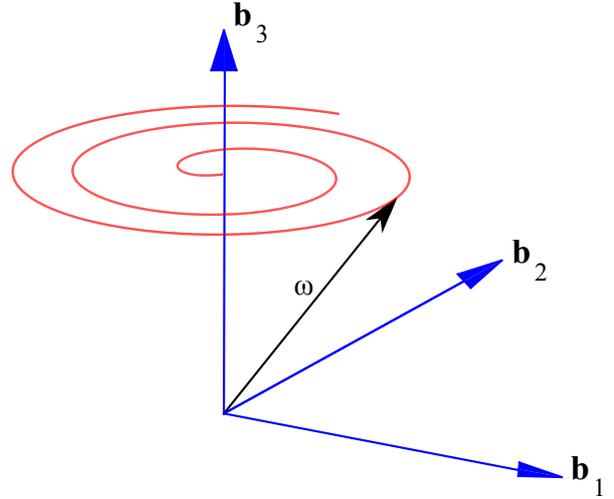}
  \caption{Uniform burn: body surface}
  \label{fig:4.5}
\end{figure}

The rotation of {\boldmath $\omega$} about ${\bf b}_3$ and ${\bf n}_h$ was discussed
in the earlier section on generic axisymmetric variable mass
systems. Figure \ref{fig:4.5} visualizes the first of these rotations for the uniformly
burning cylinder, assuming $k_3>k_1$. The angular velocity vector traces a spiral
in the transverse plane as it rotates about the symmetry axis ${\bf b}_3$ in a
counter-clockwise direction and eventually converges to the symmetry axis.
\begin{figure}[h]
  \centering\includegraphics[scale=0.8]{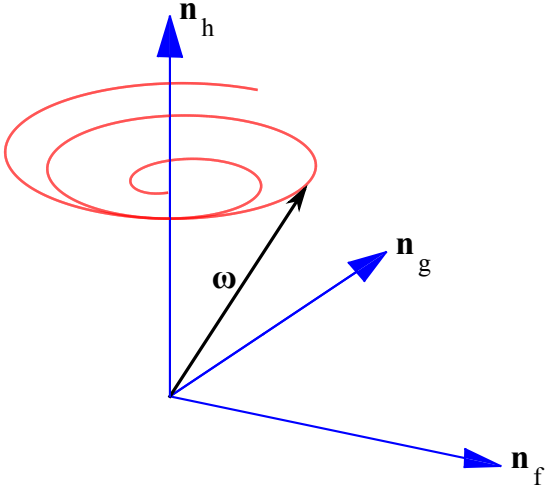}
  \caption{Uniform burn: space surface}
  \label{fig:4.6}
\end{figure}

The rotation of $\omega$ in the inertial space (i.e. about ${\bf n}_h$ since it is
inertially fixed) requires knowledge of the angle, $\beta$, in
Equation (\ref{40})
\begin{equation}
  \label{44}
  \beta = \tan^{-1}\bigg(\frac{\omega_{0}}{\omega_{30}}
  \bigg(\frac{m}{m(0)}\bigg)^{\frac{2h^2}{3k_1^2}} \bigg).
\end{equation}
This angle also decays with time for the uniformly burning cylinder, effectively reducing
the system's motion to that of simple or pure spin i.e. a motion in which the angular
velocity and angular momentum vectors are parallel. The above expressions of $\theta$, and
$\beta$ along with the angular velocity permit the construction of the space surface
as shown in Figure \ref{fig:4.6}. Note that ${\bf n}_f$, ${\bf n}_g$, and
${\bf n}_h$ represent a dextral set of unit vectors which are inertially fixed. 
\subsection{End Burn}
\begin{figure}[h]
  \centering\includegraphics[scale=.65]{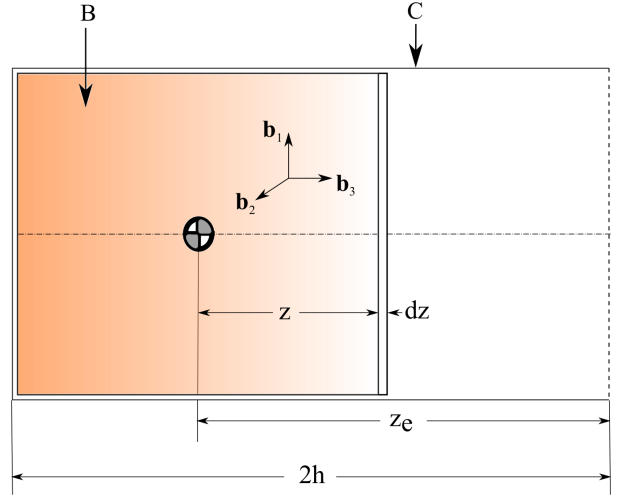}
  \caption{End burning cylinder}
  \label{fig:4.7}
\end{figure}
In this mechanism of mass variation, the cylinder burns from the circular face closest to the
exit plane towards its other face. At any instant, the unburned part retains the shape of a
right circular cylinder. Figure \ref{fig:4.7} represents a cylinder that burns from right
to left. The axial radius of gyration for such a cylinder is constant, given by
$k_3^2 = R^2/4$. The corresponding axial moment of inertia is
\begin{equation}
  \label{45}
  J = m R^2/4 \quad 
\end{equation}
As in the case of uniform burn, the spin rate can thus be determined from Equation
(\ref{18}) to be constant,
\begin{equation}
  \label{46}
  \omega_3 = \omega_{30}.
\end{equation}

The more interesting development occurs in the transverse directions. The transverse radius
of gyration is $k_1^2 = \frac{R^2}{4} + \frac{z^2}{3}$. Due to the changing length of the
cylinder, $k_1$ is not a constant.  Then, the transverse moment of inertia is
\begin{equation}
  \label{47}
        I = m(\frac{R^2}{4} + \frac{z^2}{3})
\end{equation}
If $z$ is the instantaneous half length of the unburned cylinder, then its instantaneous mass
is given by 
\begin{equation}
  \label{48}
  m = \pi R^2 (2z)
\end{equation}
The rate of mass loss is then
\begin{equation}
  \label{49}
  \dot m = 2 \pi R^2 \dot z.
\end{equation}
which can also be written as
\begin{equation}
  \label{50}
  \mathrm{d}m = 2 \pi R^2 \mathrm{d}z.
\end{equation}
From Equations (\ref{48}) and (\ref{50}), it is possible to recast Equation (\ref{33}) as
\begin{eqnarray}
  \label{51}
  \omega_{12} = \omega_0 \bigg(\frac{I(0)}{I}\bigg)
  \text{exp}\bigg[ \int_{z_0}^z
    			\frac{\bigg[z_e^2 + \frac{R^2}{4}\bigg]}
		   	{\frac{R^2}{4} + \frac{z^2}{3}}
	            	\frac{\mathrm{d}z}{z}\bigg].
\end{eqnarray}
Substituting $z_e = 2h - z$ into Equation (\ref{51}) gives
\begin{equation}
  \label{52}
  \omega_{12} = \omega_0 \bigg(\frac{I(0)}{I}\bigg)
  \text{exp}\bigg[ \int_{z_0}^z
    			\frac{\bigg[z^2 + 4h^2 -4hz + \frac{R^2}{4}\bigg]}
		   	{\frac{R^2}{4} + \frac{z^2}{3}}
	            	\frac{\mathrm{d}z}{z}\bigg]
\end{equation}
which can also be reformulated as
\begin{eqnarray}
  \label{53}
  \omega_{12} = &&\omega_0 \bigg(\frac{I(0)}{I}\bigg)
  	   \text{exp}\bigg[ 
    		\int_{z_0}^z \frac{3z - 12h}{\frac{3R^2}{4} + z^2} \mathrm{d}z \nonumber\\
	   	&&+ \bigg(12h^2 + \frac{3R^2}{4} \bigg)\int_{z_0}^z \frac{\mathrm{d}z}
		    {z\bigg(\frac{3R^2}{4} + z^2\bigg)}
	    \bigg].
\end{eqnarray}
A partial fraction expansion of the second integral in Equation (\ref{53}) then gives
\begin{eqnarray}
  \label{54}
  \omega_{12} = &&\omega_0 \bigg(\frac{I(0)}{I}\bigg)
  		\text{exp}\bigg[ 
    		\int_{z_0}^z\frac{3z - 12h}
			{\frac{3R^2}{4} + z^2} \mathrm{d}z \nonumber\\
	   &&+ \bigg(\frac{16h^2}{R^2} + 1 \bigg) \int_{z_0}^z \bigg(\frac{1}{z}
	   		- \frac{z} {\frac{3R^2}{4} + z^2}\bigg) \mathrm{d}z	
	    \bigg]
\end{eqnarray}
or, after some algebra,
\begin{eqnarray}
  \label{55}
  \omega_{12} = &&\omega_0 \bigg(\frac{I(0)}{I}\bigg)
  		\text{exp}\bigg[
    		  -\int_{z_0}^z \frac{12h}{ \frac{3R^2}{4} + z^2 } \mathrm{d}z \nonumber \\  
		 &&- \bigg(\frac{16h^2}{R^2} - 2 \bigg) 
		 	\int_{z_0}^z \frac{z}{\frac{3R^2}{4} + z^2} \mathrm{d}z\nonumber\\
	   	 &&+ \bigg(\frac{16h^2}{R^2} + 1 \bigg) \int_{z_0}^z\frac{\mathrm{d}z}{z}
       	    	\bigg].
\end{eqnarray}
The solution to the transverse rate is
\begin{eqnarray}
  \omega_{12} = \omega_0 \bigg(\frac{I(0)}{I}\bigg)
  		\text{exp}\bigg[
		  &&\frac{8 \sqrt{3}h}{R} 
		  	\tan^{-1}\bigg( \frac{2\sqrt{3}R(z-h)}{3R^2 + 4zh} \bigg) \nonumber\\
		  &&- \bigg(\frac{8h^2}{R^2} - 1 \bigg) \log\bigg( \frac{\frac{3R^2}{4} + z^2}
		  	{\frac{3R^2}{4} + h^2}\bigg) \nonumber\\
		   &&+ \bigg(\frac{16h^2}{R^2} + 1 \bigg) \log{\frac{z}{h}}
       	    	\bigg].
\end{eqnarray}
which can also be expressed as
\begin{eqnarray}
  \omega_{12} = &&\omega_0 \bigg(\frac{I(0)}{I}\bigg)
		   \bigg( \frac{\frac{3R^2}{4} + h^2}
		  {\frac{3R^2}{4} + z^2}\bigg)^{ \bigg(\frac{8h^2}{R^2} - 1 \bigg)}
		   \bigg(\frac{z}{h}\bigg)^ {\bigg(\frac{16h^2}{R^2} + 1 \bigg)} \nonumber\\
 		  &&\text{exp}\bigg[\frac{8 \sqrt{3}h}{R} 
		  	\tan^{-1}\bigg( \frac{2\sqrt{3}R(z-h)}{3R^2 + 4zh} \bigg)
       	    	\bigg].
\end{eqnarray}
The above solution can also be expressed in terms of the transverse
radius of gyration as
\begin{eqnarray}
  \label{56}
  \omega_{12} = \omega_0 &&\bigg(\frac{k_1(0)}{k_1}\bigg)^{\frac{8h^2}{R^2}}
  \bigg(\frac{z}{h}\bigg)^{\frac{16h^2}{R^2}}\nonumber\\
  &&\text{exp}\bigg[\frac{8 \sqrt{3}h}{R} 
    \tan^{-1}\bigg( \frac{2\sqrt{3}R(z-h)}{3R^2 + 4zh} \bigg)
  \bigg].
\end{eqnarray}
Equation (\ref{56}) explains that, for an end burning cylinder, the transverse rate is
regulated by 3 time varying functions: the ratio of the initial transverse radius of gyration
to its instantaneous value, the ratio of the instantaneous half length to the initial half
length, and an exponential function. The first of these functions grows as the burn
progresses but is always bounded as its denominator is never zero. The
latter two functions, however, always decay with time. Thus, irrespective of the nature
of the cylinder, the transverse rates for the end burn are always decaying; the rate
of decay is seen to be faster for prolate cylinders when compared to oblate
cylinders, which makes sense in view of Equation (\ref{56}). This has also been observed
in Reference [\citenum{ekewang2}].

\begin{figure}
  \centering\includegraphics[scale=0.7]{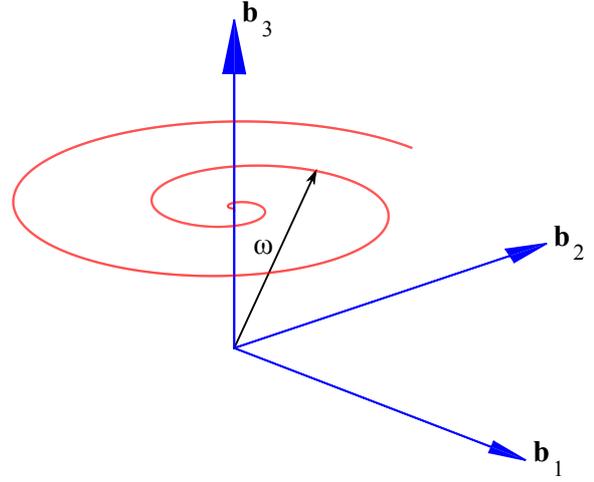}
  \caption{End burn: body surface ($J_0>I_0$)}
  \label{fig:4.8}
\end{figure}

\begin{figure}
  \centering\includegraphics[scale=0.7]{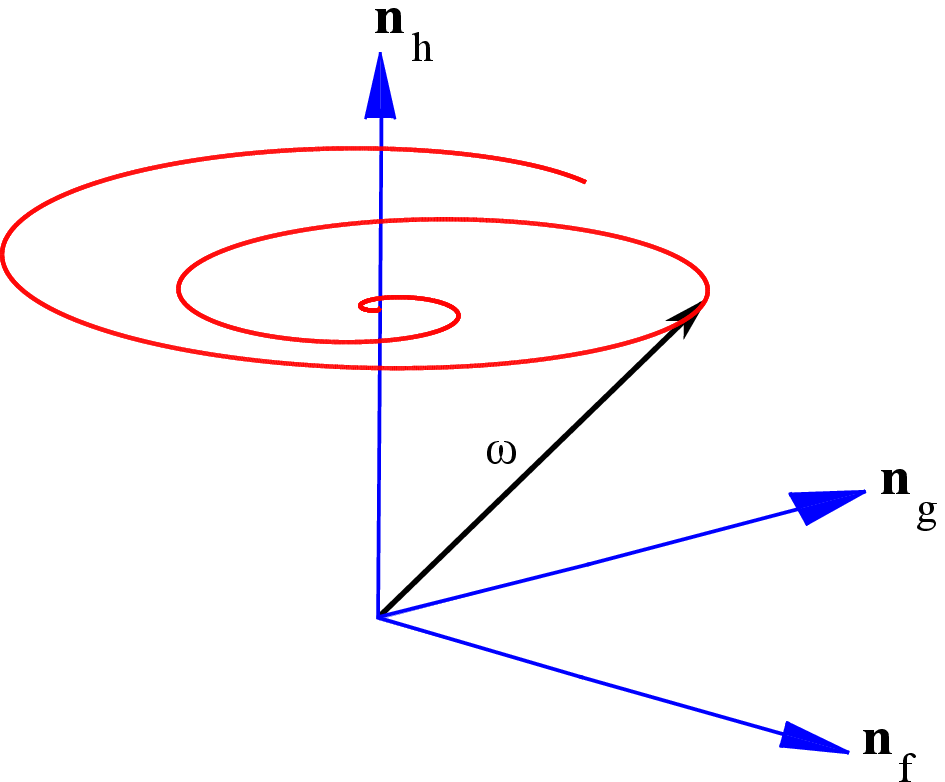}
  \caption{End burn: space surface ($J_0>I_0$)}
  \label{fig:4.9}
\end{figure}

\begin{figure}
  \centering\includegraphics[scale=0.7]{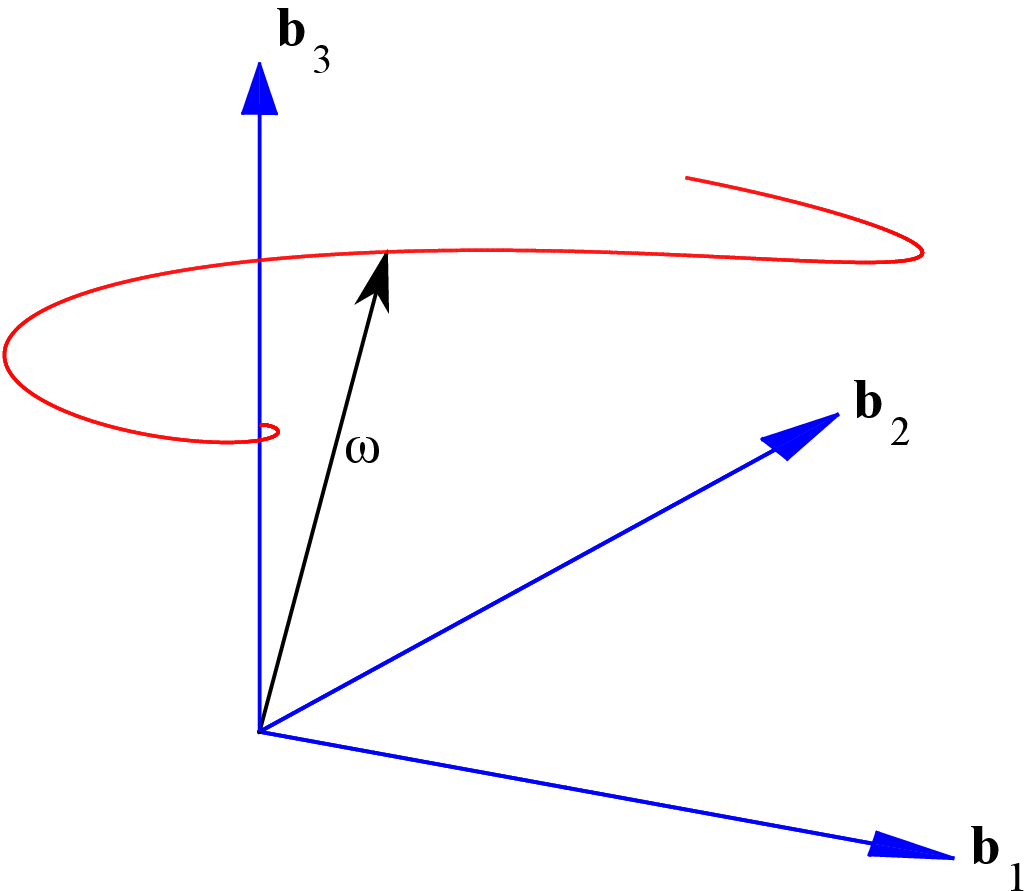}
  \caption{End burn: body surface ($J_0<I_0$)}
  \label{fig:4.10}
\end{figure}

\begin{figure}{H}
  \centering\includegraphics[scale=0.8]{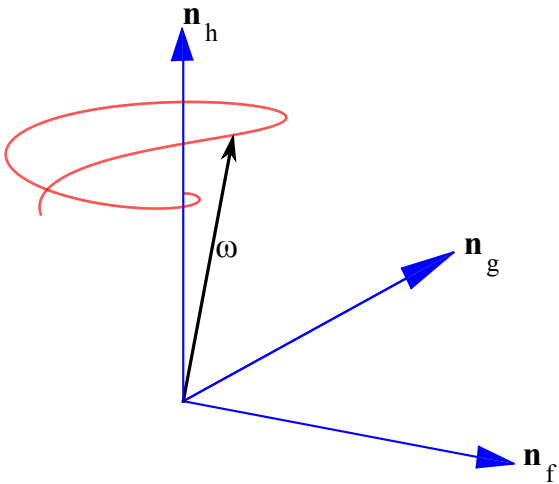}
  \caption{End burn: space surface ($J_0<I_0$)}
  \label{fig:4.11}
\end{figure}

The nutation angle for an end-burning cylinder is given by
\begin{equation}
  \label{57}
  \theta(t) = \tan^{-1}( K k_1^2 \Gamma(t) ),
\end{equation}
where $K = \omega_0/k_3^2 \omega_{30}=$ constant. As both $\Gamma(t)$ and $k_1$ are generally
decreasing with time, the nutation angle decays with time for any end burning cylinder.
In other words, jet damping is observed for such a mass-varying cylinder. Figures
\ref{fig:4.8} and \ref{fig:4.9} show the body and space surfaces for end burning cylinders
where the $J_0>I_0$ (initial $L$ = 1 m and $R$ = 0.8 m) whereas Figures
\ref{fig:4.10} and \ref{fig:4.11} are for end-burning cylinders where $J_0<I_0$
(initial $L$ = 1 m and $R$ = 0.5 m). It is interesting to note in the second case
(where $J_0<I_0$) that the angular velocity vector changes its direction of
rotation, about both ${\bf b}_3$ and ${\bf n}_h$. Initially, it rotates in a clockwise
direction and then switches to a counterclockwise direction, eventually converging to a
pure spin about ${\bf b}_3$ in the body frame and ${\bf n}_h$ in the inertial frame.
This transition point occurs when the ratio of the axial radius of gyration to the
transverse radius of gyration falls below 1 and changes the sign of
$\dot \chi$ as given by Equation (\ref{24}). In either case, the end burn is seen to
exhibit a decay in the cone angle and is thus nutationally stable.

\subsection{Radial burn}
\begin{figure}[t]
  \centering\includegraphics[scale=0.42]{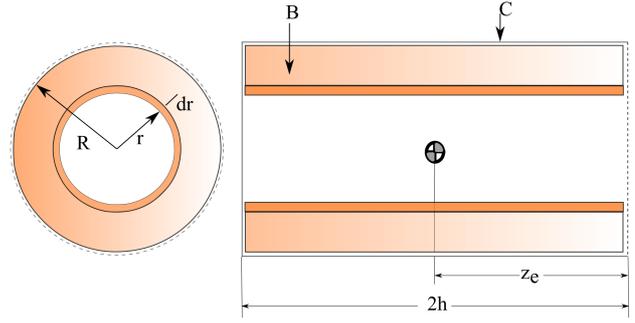}
  \caption{Radially burning cylinder}
  \label{fig:4.12}
\end{figure}

\begin{figure}[h]
  \centering\includegraphics[scale=0.8]{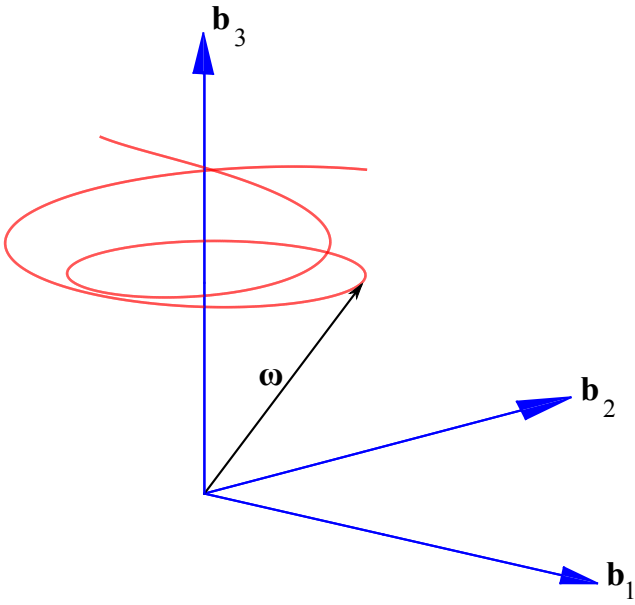}
  \caption{Radial burn: body surface}
  \label{fig:4.15}
\end{figure}

\begin{figure}[h]
  \centering\includegraphics[scale=0.8]{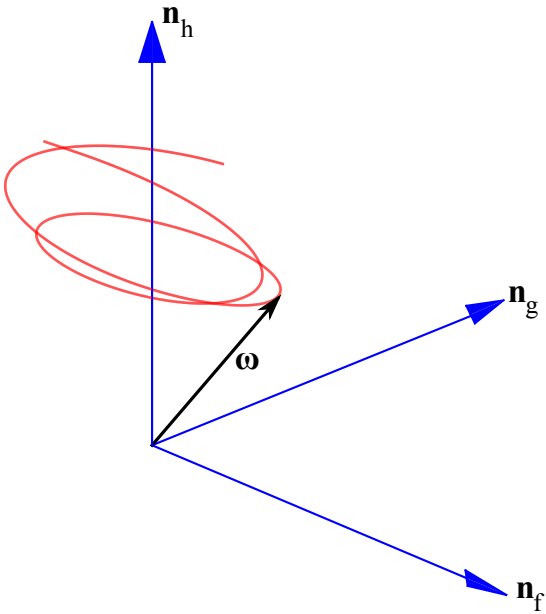}
  \caption{Radial burn: space surface}
  \label{fig:4.16}
\end{figure}

In radial burn, combustion starts along the symmetry axis and proceeds radially outwards.
Thus, at any instant, the unburned portion resembles a hollow pipe and both the axial and
transverse radius of gyration of the system are variable as shown in Figure
\ref{fig:4.12}.
For the results presented here, initial radius and length are $1$ m. The
instantaneous moment of inertia scalars for the radially burning cylinder are
\begin{equation}
  \label{58}
  I = m(\frac{R^2 + r^2}{4} + \frac{h^2}{3})
\end{equation}
and
\begin{equation}
  \label{59}
  J = m\frac{R^2 + r^2}{2}.
\end{equation}
Their corresponding time derivatives are
\begin{equation}
    \begin{aligned}
        \dot I &= \dot m(\frac{R^2}{2} + \frac{h^2}{3})\\
        \dot J &= \dot m {r^2}.
    \end{aligned}
\end{equation}
The angular speeds evaluate to
\begin{eqnarray}
  \label{60}
  &\omega_3 = \omega_{30} \frac{R^4}{(R^2 + r^2)\sqrt{(R^2 + r^2)(R^2 - r^2)}}\\
  &\omega_{12} = \omega_0 \bigg(\frac{R^2 + {4h^2}/3}{R^2 + {4h^2/3 + r^2}}
  \bigg)^\frac{(3R^2 + 16h^2/3)}{(2R^2 + 4h^2/3)}
  \bigg(\frac{R^2 - r^2}{R^2} \bigg)^\frac{(-R^2 + 8h^2/3)}{(2R^2 + 4h^2/3)}.
\end{eqnarray}
Evidently, the spin rate is not constant for a cylinder subject to radial burn.
It varies such that it appears to damp out, slowly, for about two-thirds of the burn.
Towards the end of the burn, i.e. as $r$ approaches $R$, the spin rate grows without
bounds. Analytically, this growth in spin rate has been found to begin at the
instant $r/R = 0.707$ [\citenum{ekemao3}]. The transverse angular speed, on the
other hand, can either grow, decay, or fluctuate depending on the shape of the
cylinder before the start of the burn. The corresponding body and space surfaces
for the radially burning cylinder are shown in Figures \ref{fig:4.15} and \ref{fig:4.16},
respectively, for a burn duration of $90$ s; this duration is chosen purely to
study a case where the end mass of the system approaches the mass of a payload. 

The case presented here is for $R/h=2$ and is a nutationally stable configuration; this
is in agreement with Mao and Eke's work [\citenum{ekemao3}] where it was shown that
transverse rates are unbounded for oblate radially burning cylinders with a
$R/h \geq \sqrt{8/3}$. The cone angle does in fact monotonically decrease though this
is not as evident from Figure \ref{fig:4.15} when compared to the uniform and end burns;
this is primarily due to the fact that the system's end mass is approximately 315 kg at the
end of this radial burn whereas in the other burns it is nearly zero. One approach
to verifying the nutational stability of the system is by plotting the time evolution
of $\theta$. However, the stability can also be verified from the angular speeds.
For the extreme case of oblateness represented by a radially burning flat disk
(i.e. $R>>h$), Equation (68) can be reformulated as
\begin{equation}
  \label{61}
  \omega_{12} = \omega_0 \frac{\omega_3}{\omega_{30}}
\end{equation}
by dropping the terms involving $h$ from the solution to the transverse speed. Extending
this rationale to the inertia scalars given by Equation (\ref{58}) and (\ref{59}), we
also get $I/J=1/2$. Then, the nutation angle from Equation (\ref{39}) can be expressed as
\begin{equation}
  \label{62}
  \theta = \tan^{-1}\bigg(\frac{\omega_0}{2\omega_{30}} \bigg) = \text{constant}.
\end{equation}
which is clearly bounded. In other words, even the most oblate configuration of a
radially burning cylinder results in a constant nutation angle and thus does not show
growth in the coning motion. However, unlike the preceding burns, the radial burn would
require a nutation damper system to attenuate the cone angle and bring the system into
a state of pure spin.




\section{Conclusion}
A graphical approach to study the coning motion, also known as the geometry of motion, of
a torque-free, axisymmetric variable mass system is developed. Based on this, a
general solution to the second Euler angle is also presented. The theoretical development
is seen to augment well with the work on constant mass rigid bodies. Following these
developments, the coning motion is specifically explored
for an axisymmetric cylinder with mass loss subject to three idealized burns:
the uniform, end, and radial burns. The uniform and end burning cylindrical
configurations exhibit constant spin rates and bounded decaying transverse rates; they
are seen to be nutationally stable. The radial burn generally exhibits unbounded
spin rates whereas the transverse rate is unbounded only for more oblate configurations.
More interestingly, even these oblate configurations are seen to show stable coning
motion. In relation to rocket dynamics, the results presented here show that oblate
configurations that exhibit a ``misbehavior'' in their transverse rates could
still be nutationally stable. The difference between the radial burn from the other burns
examined is that the system does not always approach a state of pure spin. This work
recommends that examining the transverse speed alone may be
insufficient in determining the stability of a freely rotating variable mass system. 

The rotational behavior of a free axisymmetric variable mass system is remarkably
different from its constant mass counterpart. For a free axisymmetric system with mass
variation, spin rates are functions of time that either grow, decay, or remain
constant whereas their transverse speeds can either grow or decay; on the other hand,
the spin and transverse speeds of constant mass systems are of constant magnitude. As
the inertia properties in the latter case are also constant, the corresponding nutation
angle (or second Euler angle) is also unchanging which leads to the well known `cones of
motion' of the angular velocity. In the case of variable mass systems, the nutation
angle is not as straightforward to generally characterize on account of the
time-varying nature of both the angular velocity and its central inertia scalars. Thus,
the results of the analyses presented clearly indicate that mass variation majorly
impacts the dynamics of freely spinning systems.

\section*{Acknowledgments}
The author would like to acknowledge the University of California Davis'
Drake Fellowship award and the Mechanical and Aerospace Engineering Department's
N. \& M. Sarigul–Klijn Space Engineering/Flight Research award that supported this work.

\section*{Compliance with ethical standards}
{\bf Conflict of interest} The authors declare that they have no conflict of interest.

\end{document}